\documentclass[aps,prb,floatfix,twocolumn,superscriptaddress,showpacs,epsf]{revtex4-1}

\usepackage[USenglish]{babel}
\usepackage{amsmath,amssymb,amsfonts}
\usepackage{epstopdf}
\usepackage{epsfig} 
\usepackage{graphicx}
\usepackage{subfigure}
\usepackage{import}
\usepackage{color}
\usepackage{url}

\allowdisplaybreaks

\usepackage[colorlinks=true,linkcolor=blue,citecolor=red]{hyperref}

\newcommand{\ca}{c^{\phantom{\dagger}}}
\newcommand{\cc}{c^\dagger}
\newcommand{\da}{d^{\phantom{\dagger}}}
\newcommand{\dc}{d^\dagger}

\newcommand{\be}{\begin{equation}}
\newcommand{\ee}{\end{equation}}
\newcommand{\bea}{\begin{eqnarray}}
\newcommand{\eea}{\end{eqnarray}}
\newcommand{\ba}{\begin{eqnarray*}}
\newcommand{\ea}{\end{eqnarray*}}

\newcommand{\up}{\uparrow}
\newcommand{\down}{\downarrow}

\newcommand{\ket}[1]{|{#1}\rangle}
\newcommand{\bra}[1]{\left\langle{#1}\right|}
\newcommand{\braket}[3]{\langle{#1}| {#2} |{#3} \rangle}

\newcommand{\quave}[1]{\langle {#1} \rangle}

\newcommand{\kc}{K$_3$C$_{60}~$}

\def\eg{\mbox{\it e.g.\ }}

\begin{document}

\title{Non-equilibrium Superconductivity in driven alkali-doped fullerides}
\author{Giacomo Mazza}
\affiliation{Centre de Physique Th\'eorique, \'Ecole Polytechnique, CNRS, Universit\'e Paris-Saclay, 91128 Palaiseau, France}
\affiliation{Coll\`ege de France, 11 place Marcelin Berthelot, 75005 Paris, France}
\author{Antoine Georges}
\affiliation{Coll\`ege de France, 11 place Marcelin Berthelot, 75005 Paris, France}
\affiliation{Centre de Physique Th\'eorique, \'Ecole Polytechnique, CNRS, Universit\'e Paris-Saclay, 91128 Palaiseau, France}
\affiliation{Department of Quantum Matter Physics, University of Geneva, 24 Quai Ernest-Ansermet, 1211 Geneva 4, Switzerland}


\begin{abstract}
We investigate the formation of non-equilibrium superconducting states in driven alkali-doped fullerides A$_3$C$_{60}$. 
Within a minimal three-orbital model for the superconductivity of
these materials, it was recently demonstrated theoretically that an orbital-dependent imbalance of the 
interactions leads to an enhancement of superconductivity at
equilibrium [M.~Kim {\it et al.} Phys. Rev. B 94, 155152 (2016)].
We investigate the dynamical response to a time periodic modulation of this interaction imbalance, 
and show that it leads to the formation of a transient superconducting state which survives 
much beyond the equilibrium critical temperature $T_c$.
For a specific range of frequencies, we find that the driving reduces
superconductivity when applied to a superconducting state below
$T_c$,  while still inducing a superconducting state when the initial temperature is larger than $T_c$.
These findings reinforce the relevance of the interaction-imbalance mechanism as a possible explanation of the recent 
experimental observation of light-induced superconductivity in alkali-doped fullerenes.
\end{abstract}

\maketitle
\section{Introduction}
The optical stimulation of solids by means of strong light
pulses has opened new routes for the investigation of collective phenomena in 
quantum materials~\cite{giannetti_review}.  A fascinating one consists in inducing superconductivity (SC) 
beyond the limits where it can be stabilized at equilibrium, which are set \eg by 
temperature, external pressure or doping concentration. 
A series of experiments in different compounds revealed light-induced modifications of the electronic properties 
suggestive of the formation of a transient  superconducting state 
extending above the equilibrium critical
temperature $T_c$~\cite{fausti_science,kaiser_ybco,hu_ybco,non_linear_phononics_ybco,mitrano_k3c60}. 
Recently, the remarkable observation of a superconducting-like response above $T_c$
in the molecular compound \kc\cite{mitrano_k3c60} 
enlarged this experimental panorama and raised new questions about the 
possible mechanisms leading to transient SC above $T_c$.

The possibility of a light-stimulated superconducting phase extending above the 
equilibrium critical temperature has been originally proposed by Eliashberg~\cite{eliashberg} 
who considered the quasiparticle redistribution induced by a laser excitation with
frequencies below the equilibrium superconducting gap.
The above experiments are far beyond this limit with excitation frequencies 
much larger than the superconducting gap, thus requiring
the investigation of alternative mechanisms for light-induced superconductivity.

In the case of the \kc the laser frequencies for which
the transient response is observed are close to the frequencies of four 
intra-molecular phonons, the $T_{1u}$ modes with frequencies in the mid-infrared range 
$60 \lesssim \omega_{T_{1u}} \lesssim  180~\text{meV} $. The effect 
disappears for much larger excitation frequencies. This suggests 
that the observed
effect might be related to the light-induced excitation of these phonon modes.~\cite{mitrano_k3c60,minjae_dU}

From the theoretical point of view, various mechanisms have been so far investigated
such as the non-linear excitations of phononic modes~\cite{non_linear_phononics_ybco,murakami_eph,knap_eph}
and their coupling to the electronic density~\cite{kennes_squeezing_phonons,sentef_enhanced_eph}
or the effective slowing-down of the electronic motion~\cite{sentef_eph,coulthard_drivenU}.
While all these mechanisms lead to an increase of the superconducting coupling
which is expected to provide a source of transient SC in a broad class of superconductors,
\kc appears as a peculiar case.
Indeed, the absence of any enhancement or even the suppression of SC below $T_c$ 
reported in Ref.~\onlinecite{mitrano_k3c60}, together with the appearance of a 
transient response above $T_c$ is not fully understood within an effective SC coupling enhancement.
Furthermore, SC in alkali-doped fullerides is strongly affected by the non-trivial interplay between pairing, 
electronic correlations and orbital degrees of 
freedom~\cite{massimo_science2002,capone_RMP,yusuke2015,gunnarsson_jt_phonons,gunnarsson_RMP},
requiring the investigation of the mechanisms for transient SC
within a proper theoretical framework taking this interplay into account. 

A first step in this direction has been recently taken by Kim et al.~\cite{minjae_dU},
working in the framework of the low-energy electronic description of fullerides based on the Jahn-Teller 
induced inversion of the effective Hund's coupling~\cite{capone_RMP}, which provides one of the most successful
description of the unconventional superconducting properties of these materials~\cite{yusuke2015}.
These authors demonstrated  that a specific
orbital-dependent perturbation of the on-site repulsive interactions does enhance SC at equilibrium.
Such a perturbation was motivated by the possible modulations
of the electronic interactions that result from the excitation
of a local phononic mode, as already demonstrated for other correlated
organic compounds~\cite{kaiser_Umodulation,singla_thz_Umodulation}.
Furthermore, a first principle calculation~\cite{minjae_dU} for \kc 
revealed that the favorable perturbation is indeed induced, 
under the assumption that light excites the $T_{1u}$ mode~\cite{mitrano_k3c60}, 
as a result of the structural and electronic changes associated with this excitation.

This proposal relies entirely on equilibrium considerations however, and this raises the outstanding question 
of the relevance of this mechanism to the non-equilibrium response of the system and to the transient light-induced SC.  
In this work we address this question.
We investigate the non-equilibrium dynamics induced by the time- and orbital- dependent modulation of the electron-electron repulsion. 
We show that this results in a transient superconducting state, 
which can be induced when the system is initially well above its equilibrium critical temperature. 
The properties of this transient state can be very dramatically
different from the equilibrium expectation and depend on the frequency of the modulation.
In particular, we uncover a regime of frequency in which the modulation leads 
to the reduction of SC below $T_c$ and to the creation of SC above
$T_c$.

In the following we will first introduce the model and the
non-equilibrium perturbation considered in this work together with the
method used to describe the non-equilibrium dynamics. After presenting our
results we will discuss possible implication for the description of
the experimental observations.

\section{Model}
\label{sec:model}
The minimal description of strongly correlated superconductivity in
alkali-doped fullerenes is given by the following three-bands model~\cite{capone_RMP}
arising from the $t_{1u}$ LUMO states of the C$_{60}$ molecule
half-filled with electrons donated by the alkali atoms
\begin{equation}
\mathcal{H} = \sum_{\mathbf{k} \sigma} \sum_{a=1}^3
\epsilon(\mathbf{k}) \cc_{\mathbf{k} a \sigma} \ca_{\mathbf{k} a
  \sigma} + \sum_{i} \mathcal{H}_{loc}(i),
\label{eq:Hamiltonian}
\end{equation}
where the local Hamiltonian $\mathcal{H}_{loc}(i)$ is of the Kanamori
type~\cite{kanamori} and takes into account intra- and inter- orbital electron electron repulsion,
spin-flip and pair hopping terms, with a negative (inverted) Hund's coupling $J_H$
resulting from the competition between the Hund's coupling and the Jahn-Teller
intramolecular interactions~\cite{capone_RMP}. The explicit expression
of the local Hamiltonian reads
\begin{equation}
  \begin{split}
  \mathcal{H}_{loc} &= \sum_{\alpha} U_{\alpha} n_{i\alpha \up}
  n_{i\alpha \down} \\
  &\phantom{=}+ (U-2 J_H) \sum_{\alpha \ne \alpha'} n_{i\alpha \up}
  n_{i\alpha' \down} \\ 
  &\phantom{=}+  (U-3 J_H) \sum_{\alpha < \alpha' \sigma} n_{i\alpha \sigma}
  n_{i\alpha' \sigma} \\
  &\phantom{=}+ J_H \sum_{\alpha\ne\alpha'} \cc_{i \alpha \up} \cc_{i \alpha' \down}
  \ca_{i \alpha \down} \ca_{i \alpha' \up} \\
  &\phantom{=}+  J_H \sum_{\alpha\ne\alpha'} \cc_{i \alpha \up} \cc_{i \alpha \down} \ca_{i \alpha' \down} \ca_{i \alpha' \up},
  \end{split}
  \label{eq:Hloc}
\end{equation}
where $\alpha,\alpha'$ and $\sigma$ indices indicates orbital and spin degrees
of freedom respectively.

At equilibrium the interaction terms on each orbital are degenerate
$U_{\alpha}=U$. 
The phonon excitation induces the modification of the local
interactions energies $U_{\alpha}$ due to the coupling between the local electronic
configurations and the coordinate of the displaced phononic mode along
a given direction $q(t) = A \sin \Omega t$.
In general, this is due to the fact that, for an odd parity mode,
such as $T_{1u}$,
at the lowest order the displaced phononic coordinate couples
quadratically with the local double occupied
states.~\cite{kaiser_Umodulation,singla_thz_Umodulation} For a single
band case this leads to an additional term in the local Hamiltonian
\begin{equation}
\mathcal{H}_{e-ph} \propto q(t)^2 n_{\up} n_{\down}, 
\label{eq:qsquare_single}
\end{equation}
meaning the oscillation of the local interaction with a frequency
$2\Omega$ around a renormalized value due to the fact that square of
the mode displacement has a finite average $\quave{q(t)^2} \ne 0$. 
In the multi-band case, neglecting contributions coming from the
coupling between  different orbitals electronic configurations,
Eq.~(\ref{eq:qsquare_single}) is generalized to 
\begin{equation}
  \begin{split}
    \mathcal{H}_{e-ph} &= \sum_{\alpha} C_{\alpha} q(t)^2 n_{\alpha \up} n_{\alpha \down}\\
    & \equiv \sum_{\alpha} \Delta U_{\alpha} \left[1-\cos 2 \Omega t \right]
  n_{\alpha \up} n_{\alpha \down},
  \end{split}
  \label{eq:qsquare_multi}
\end{equation}
where the coefficients $\Delta U_{\alpha}$ are specific properties of the
phononic mode and of the direction of excitation, determined by the
light pulse polarization. 
Here we take advantage of the first principle results of
Ref.~\onlinecite{minjae_dU} showing that the displacement
of the normal coordinate of the $T_{1u}$ mode along a given direction 
leads to the removal of the orbital degeneracy between the $U_{\alpha}$, 
leading to two orbitals with smaller interaction with respect to the third one.
We insert this result in Eq.~(\ref{eq:qsquare_multi}) by considering
$\Delta U = -\left[ \delta U,\delta U,0 \right]$ with $\delta U>0$,
so that the intra-orbital interaction terms in Eq.~(\ref{eq:Hloc})
become
\begin{equation}
  U_{x,y}(t) = U - r(t) \frac{\delta U}{2}\, \left(1 -\cos 2 \Omega t\right) ; \quad
  U_z(t) = U,
  \label{eq:neqU}
\end{equation}
where $r(t)$ is a smooth ramping function, defined as
$r(t) = 1/2-3/4\cos \pi t/\tau + 1/4 \cos^3 \pi t/\tau $  for $t<\tau$ and
$r(t)=1$ for $t\geq \tau$, which phenomenologically takes into account the
time  $\tau$ during which the modulation of the $U$ is switched on.

\begin{figure}
  \includegraphics[width=1.\linewidth]{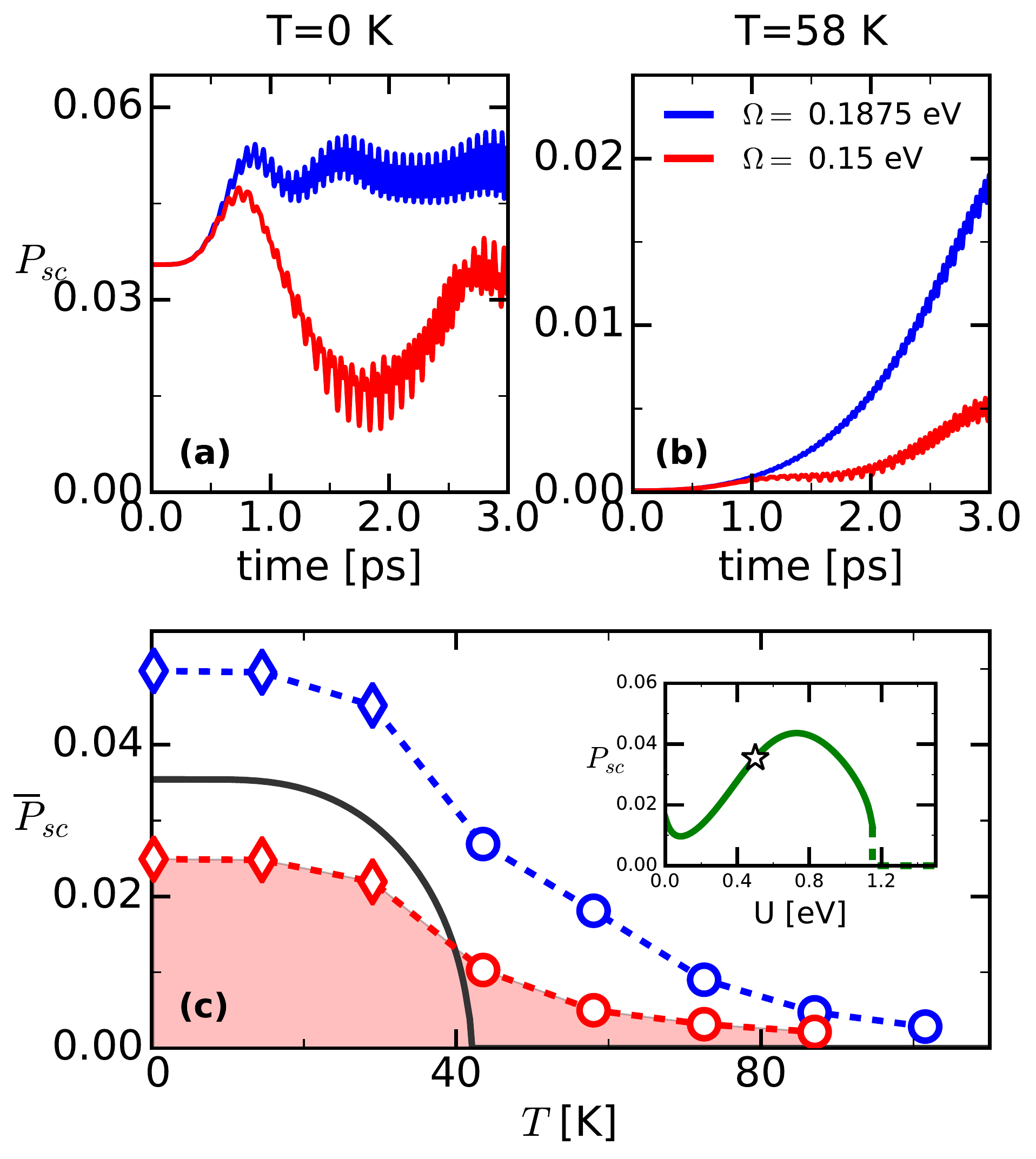}
  \caption
      {
        Panels (a)-(b):
        Dynamics of the global order parameter for two driving frequencies
        $\Omega=0.1875~\text{eV}$ (blue lines) and $\Omega=0.15~\text{eV}$
        (red lines) at zero temperature (a) and  $T = 58~\text{K} > T_{c} \simeq 42~\text{K}$ 
        (b). Panel (c): Transient order parameter as a function of temperature. 
        Color code as in panel (a).
        Shaded area highlights the regime for which SC is suppressed below $T_c$ and
        created above.
        For $T < T_c$ (diamonds) where an almost steady value is reached during the
        dynamics we extract this value taking a time average. For $T > T_c$ (circles) we take 
        the value of the switched order parameter at $t=3~\text{ps}$.
        Dashed lines are guides to the eye.
        Inset: Zero-temperature order parameter as a function of $U$.
        The star indicates the values of parameters considered in this work.
      }
      \label{fig:above_tc_neqSC}
\end{figure}

We implemented and use the time-dependent Gutzwiller approximation (tdGA)~\cite{schiro2010,michele_review} 
extended to the multi-band superconducting  case ~\cite{hugo_lanata,spin_dynamics_t2g,giacomo_michele_2012}. 
The method is based on the variational ansatz for the time evolved state 
\begin{equation}
  \left| \Psi(t) \right\rangle \simeq \prod_i \mathcal{P}_i (t)
  \left| \Psi_0(t) \right\rangle,
  \label{eq:gz_ansatz}
\end{equation}
where $\left| \Psi_0(t) \right\rangle$ is an uncorrelated wavefunction
describing the coherent quasiparticle dynamics and $\mathcal{P}(t)$ is a
projector onto the local Hilbert spaces giving the weights of the
local atomic multiplets. The dynamics of both quantities are determined
via the time-dependent variational principle $\delta \int \langle
\Psi(t) | i \partial_t - H | \Psi(t) \rangle = 0 $.
At equilibrium the variational ansatz Eq.~(\ref{eq:gz_ansatz}) is equivalent
to the rotationally invariant slave bosons technique~\cite{slave_bosons_antoine}
which has been already successfully used to describe equilibrium
strongly correlated SC in the present model~\cite{massimo_slave_bosons_scs}.

The method is extended to the finite-temperature case by the introduction of a 
time dependent variational density matrix~\cite{michele_gzT,lanata_gzT}
\begin{equation}
  \rho(t) = \mathcal{P}(t) \rho_*(t) \mathcal{P}(t)^{\dagger}
  \label{eq:gz_ansatzT}
\end{equation}
where the projector $\mathcal{P}(t)$ has the same definition as
in~(\ref{eq:gz_ansatz}) and $\rho_*(t) = \sum_{n} p_n \ket{\Psi_n(t)}
\bra{\Psi_n(t)}$ is the density matrix corresponding to
a complete set of uncorrelated  states $\ket{\Psi_n(t)}$ and a distribution $p_n$.
The dynamical equations for (\ref{eq:gz_ansatzT}) are obtained by applying the
finite temperature generalization of the Dirac-Frenkel
variational principle~\cite{frenkel_dirac_T}. 
They are solved numerically, with an initial condition corresponding to 
the equilibrium thermal state.
Details about the methods are reported in the Appendix.

\section{Results}

In the following we will consider a semicircular density of states
with a bandwidth $W=0.5$~eV and take $U=0.5~\text{eV}$
and $J_H=-0.02~\text{eV}$.
In the inset of Fig.~\ref{fig:above_tc_neqSC}(c) we show that
at equilibrium this corresponds to a superconductor on the weak correlation side
of the superconducting dome determined by the electron-electron repulsion
$U$ in the model Eq.~(\ref{eq:Hamiltonian}).
This is consistent with \eg the pressure dependence of $T_c$
observed experimentally for \kc~\cite{zadike_2015}.
We take the modulation frequency $\Omega$ as 
an adjustable parameter in a range reasonably
including the typical frequencies of $T_{1u}$ modes
and we fix $\delta U/U =0.1$.
We choose a ramping time $\tau=0.9~\text{ps}$. The following results do not
depend qualitatively on this choice.

In Fig.~\ref{fig:above_tc_neqSC}(a)-(b), we plot the dynamics of
the orbital averaged amplitude of the order parameter
$P_{sc}=\sum_{\alpha}P_{sc}^{\alpha}/3$, $P_{sc}^{\alpha} = |\quave{\cc_{\alpha,\up} \cc_{\alpha,\down}}|$,
for two different driving frequencies and at two temperatures below and above 
the equilibrium critical temperature $T_c \simeq 42~\text{K}$ 
($T=0~$K in panel (a) and $T=58~$K in panel (b)). 
A tiny symmetry breaking field is introduced for $T>T_c$ to allow SC to develop.
\footnote{The time needed for a finite order parameter to develop depends 
on this choise and, therefore, it is not related to any realistic time scale.
On the other hand, the long time value of the order parameter is independent
on this choice.}

For the larger driving frequency  ($\Omega=0.1875~\text{eV}$), we observe the 
increase of the superconducting order parameter at zero temperature and the formation
of a finite order parameter for $T>T_c$. 
This establishes that the SC-enhancement mechanism based on the imbalance of $U$ 
does apply out of equilibrium. 
In particular, the formation of a finite order parameter above $T_c$ signals 
that the initial normal metal becomes an unstable state due to the increase 
of the critical temperature induced by the average perturbation.
Both enhancement of the order parameter below $T_c$ and its formation
for $T>T_c$ are expected from the equilibrium
predictions of the average interaction imbalance
and it is due to the energetic stabilization of the local multiplets 
with a singlet pair in the $x$ or $y$ orbitals~\cite{minjae_dU}.

We notice that, in principle, the continuous modulation of the interactions might lead to 
a continuos energy absorption inside the system which would eventually destroy the transient
superconducting state. However, the persistence of the superconducting state as long as the 
interaction modulation is active suggests that no sizable energy absorption occours in
this case.  In the following we show that this strongly depends on the  frequency of 
the interaction modulation.

The strong deviation from the above equilibrium expectations is observed when
the driving frequency is lowered to $\Omega=0.15~\text{eV}$. 
Starting from the superconducting state, the order parameter undergoes a decrease instead of the expected increase.
On the other hand, a finite order parameter is  still established above 
the critical temperature, though its amplitude is smaller than the one established for
$\Omega=0.1875~\text{eV}$.

In panel (c) we compare the transient order parameter as a function
of the initial temperature to the equilibrium one. The former is
extracted as a time average over a time interval  $\Delta t = t_{max}-t_{min}$, 
$\overline{P}_{sc} = \int_{t_{min}}^{t_{max}} d\tau P_{sc}(\tau)/\Delta t$,
for $T=0~$K while we estimate it for $T>T_c$ from the value it takes at
$t=3~\text{ps}$.
For $\Omega=0.1875~\text{eV}$, the non-equilibrium perturbation
enhances SC for all temperatures, up to about $T \lesssim~100~\text{K}$, way above 
equilibrium $T_c$. On the contrary, the $\Omega=0.15~\text{eV}$ case shows a remarkable
suppression of SC for $T<T_c$ and the formation of SC up to temperatures
slightly below the previous case $T\lesssim~90~\text{K}$.

\begin{figure}[tb]
  \includegraphics[width=1.\linewidth]{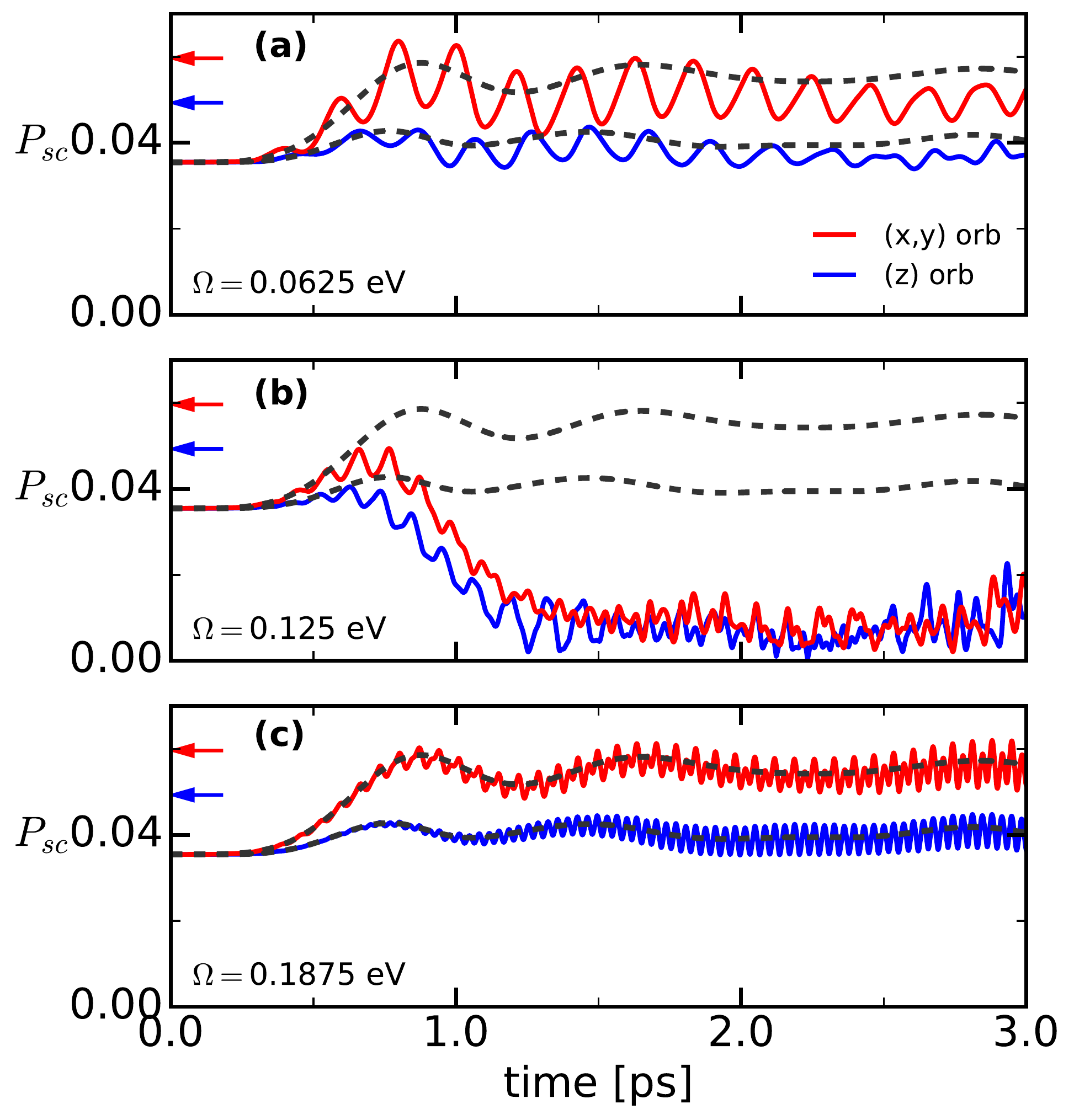}
  \caption
  {
    Zero-temperature dynamics of the order parameter
    amplitude for $x,y-$orbitals (red lines) and $z-$orbital
    (blue lines). 
    Driving frequencies $\Omega=0.0625~\text{eV}$ (a),
    $\Omega=0.125~\text{eV}$ (b) and $\Omega=0.1875~\text{eV}$ (c).
    Dashed lines: unmodulated dynamics after the switching 
    of a constant imbalance of $U$ equal to the average of the
    periodic modulation (see text). The arrows represent the
    expected equilibrium order parameters corresponding to the average interaction imbalance.
  }
  \label{fig:sc_dynamics0}
\end{figure}

The above results show that at the frequency $\Omega =
0.15~\text{eV}$ the dynamical modulation of the inteaction leads to
the suppression of the order parameter with respect to the value expected
from the sole interaction imbalance. This strongly suggests that some
energy absorption due to the continous modulation of the interaction
occurs at this value of the driving frequency.  
In order to obtain insights into this, we study the orbital resolved dynamics
of the zero-temperature superconductivity at three increasing frequencies
(Fig.~\ref{fig:sc_dynamics0}).
We compare it to the time evolution obtained by an imbalance
of $U$ equal to the average of the modulated case Eq.~(\ref{eq:neqU})
switched on during the same ramp time $U_{x,y}(t) = U -r(t)\delta U /2$,
hereafter called \emph{unmodulated dynamics}.

The imbalance of $U$ lifts the orbital degeneracy between the 
components of the superconducting order parameter $P^{\alpha}_{sc}$.
For the slowest driving frequency ($\Omega = 0.0625~\text{eV}$ in
panel (a)) the order parameter components display an oscillating
behaviour with a fast component $\omega = 2 \Omega$ which reflects the
periodical modulation of $U_{\alpha}$ and a slower component related
to the amplitude of the superconducting gap. Only the slow component
is retained in the case of the unmodulated dynamics (dashed lines).

In both cases a global enhancement of the order parameter is
observed, though smaller with respect to what is expected
from the equilibrium imbalanced case (see arrows in
Fig.~\ref{fig:sc_dynamics0}). 
This is understood from the fact that the average interaction
imbalance is switched on in a finite time, wheares the equilibrium
limit is expected to be recovered for an infinitely slow switching.

We notice that the modulated and unmodulated dynamics tend to separate
at long times ($t \gtrsim 1.5~\text{ps}$) where a larger order
parameter is established in the latter case. 
This suggests that the effect of  the continous interaction modulation 
induces some energy absorption inside the system, leading to a slow decrease
of the order parameter at long times.
As already anticipated this effect is strongly dependent on the 
modulation frequency and it is almost absent at larger driving frequency   
($\Omega = 0.1875~\text{eV}$ in panel (c)), where
 the modulated and unmodulated dynamics become almost equivalent, the
 only difference being  the fast oscillations of small amplitude in
 the modulated case.  

A dramatic effect of the dynamical modulation occurs at intermediate
frequencies. This is clearly seen on panel (b) of Fig. ~\ref{fig:sc_dynamics0} for the 
driving frequency $\Omega = 0.125~\text{eV}$ where the periodic modulation 
leads, in sharp contrast to the average interaction imbalance, to an almost
complete suppression of the superconducting order already for times of
the order of the ramping $\tau$.

\begin{figure}[tb]
  \includegraphics[width=1.\linewidth]{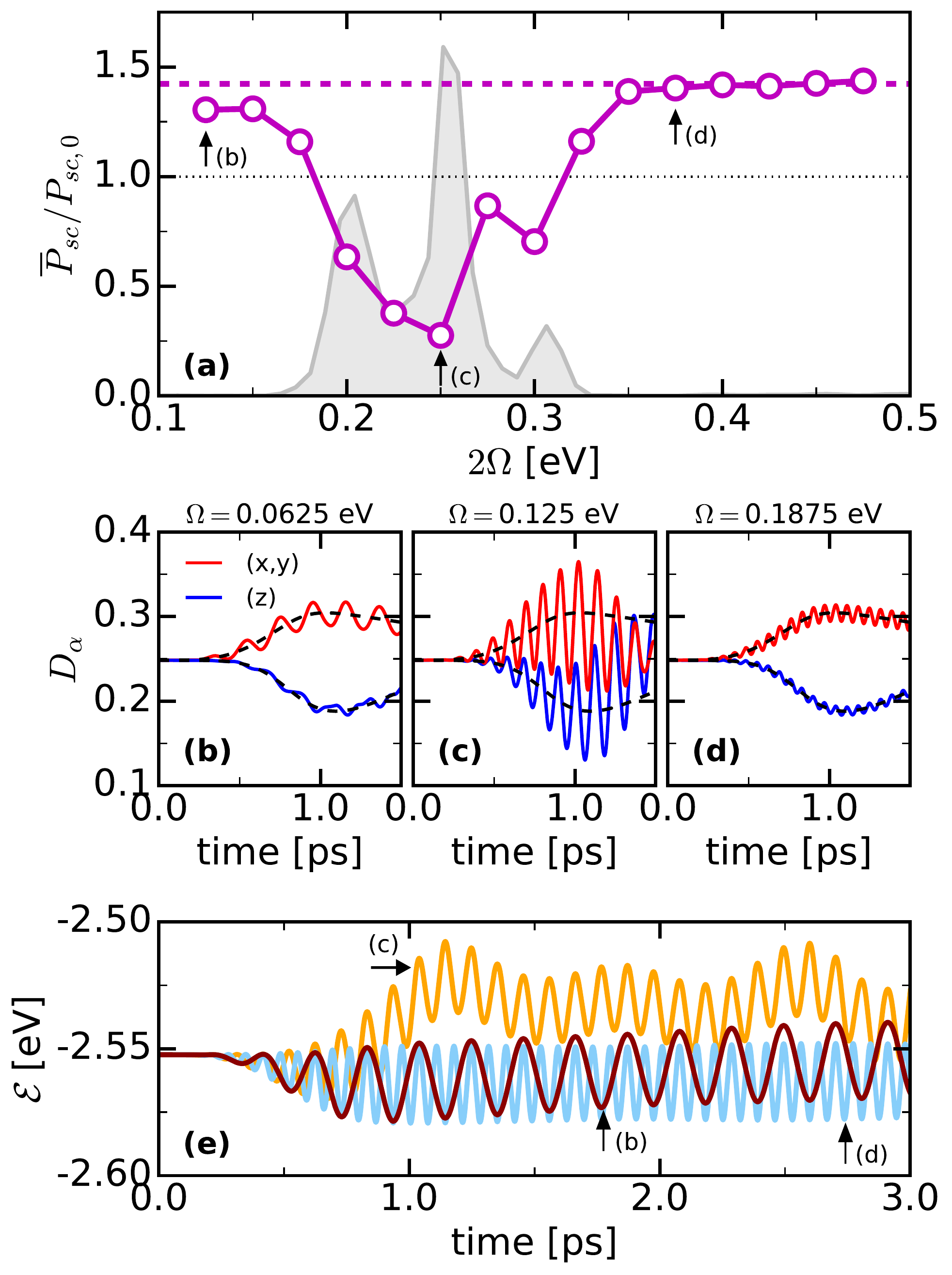}
  \caption
  {
    Panel (a): Stationary zero-temperature superconducting order parameter 
    normalized by the equilibrium value as a function of the driving frequency. 
    Dashed lines: stationary order parameter for the constant $U$-imbalance dynamics.     
    Arrows show the transient states considered in Fig.~\ref{fig:sc_dynamics0} and 
    in panels (b)-(d). Dotted line represents the equilibrium value.
    The shaded area represents the spectrum of excitations of the non-equilibrium
    doublons (see text).
    Panels (b)-(d): Dynamics of the number of average number of intra-orbital double occupations
    for the driving frequencies $\Omega=0.0625~\text{eV}$ (b),
    $\Omega=0.125~\text{eV}$ (c) and $\Omega=0.1875~\text{eV}$ (d)
    and the constant $U$-imbalance (dashed lines).
    Panel (e): Dynamics of the internal energies for the three frequencies in panels (b)-(d) (see arrows).
  }
  \label{fig:sc_vs_freq}
\end{figure}

The non-monotonic response of the transient state
to the frequency of the interaction modulation is summarized in Fig.~\ref{fig:sc_vs_freq},
where we report the existence of a frequency range 
$ 0.09~\text{eV} \lesssim \Omega \lesssim 0.16~\text{eV} $ in which SC is suppressed instead of enhanced. 
This frequency range is comprised between a lower frequency region in which the
enhancement of SC is smaller than the case of the unmodulated dynamics
and  a higher frequency region in which the perturbation
becomes completely anti-adiabatic with respect to the periodic change of $U$
and thus coincides with the unmodulated one.

We trace back the origin of the behaviour observed in the 
different regimes to the doublons excitations induced
by the periodic modulation  of $U$.
Panels (b)-(d) of Fig.~\ref{fig:sc_vs_freq} display the dynamics
of the double occupancies on each orbital $D_{\alpha} = \quave{n_{\alpha \up}  n_{\alpha \down}} $
for  the three frequencies representative of the different regimes.
Due to the asymmetric value of the interaction the number of pairs
is enhanced in the $(x,y)$ orbital and lowered on the $(z)$ one.
However the dynamics is markedly different in the different regimes of frequency.
At small and large frequencies (panel (b) and (d)), the dynamics closely
follows the unmodulated one with superimposed $2\Omega$ oscillations
indicating the creation of double occupancies on top of the
orbitally imbalanced populations of non-equilibrium doublons.
Such process becomes resonant in the intermediate frequency regime (panel (c))
where the strong amplification of the doublons oscillations is 
observed.
As shown by the dynamics of the  system's internal energy $\mathcal{E}(t)=\quave{\mathcal{H}}$
 (panel (e)), such a large amount of excitations leads 
to a sizable energy absorption which suppresses SC
with respect to what is expected for the unmodulated dynamics
(dashed line in panel (a)).

This shows that the energy absorption induced by the modulation
frequency is responsible for the superconductivity suppression in the
frequency region $0.09 \lesssim \Omega
\lesssim 0.16$ eV.
We find that the origin of such  behaviour is the
resonance between   the modulation frequency $2\Omega$ and the
spectrum of the non-equilibrium 
excitations of the orbitally imbalanced populations of doublons induced
by the asymmetric interaction (shaded area in panel (a)).
Such spectrum is extracted from the frequency spectrum $\mathcal{F}_D$
of the dynamics of doublons following the sudden switch of a fixed $U-$imbalance,
$U_{x,y}(t)=U-\theta(t)\delta U$~\cite{extended_dynamic_mott}.
We compute $\mathcal{F}_D$ at fixed $\delta U$  and then integrate
over a window equal to the amplitude of the interaction imbalance considered 
in the modulated dynamics $0 < \delta U < 0.05~\text{eV}$. 
The resulting spectrum has a broad three peaks structure which
exactly matches the frequency region for which SC is suppressed.

An investigation of the structure of the spectrum revealed that the
two side-bands mainly depend on the value of $J_H$ and 
disappear for $J_H=0$ indicating processes 
of inter-orbital origin. On the other hand, the central
peak weakly depends on $J_H$ and decreases with $U$ (not shown)
suggesting processes within the renormalized quasi-particle bandwidth as
expected by the fact that the number of excited doublons is in a small
quench regime ($\delta U/U = 0.1$).

We finally observe that, as already anticipated in the discussion of
Fig.~\ref{fig:sc_dynamics0}, away from the resonance a small energy absorption
is present for the slow driving case, whereas it is 
almost negligible for the fast one. This reflects in the mismatch between the 
modulated and the unmodulated dynamics for frequencies smaller than the resonance.

\section{Discussion}

The above results show that the dynamical modulation of the interaction
that can be induced by the excitation of a molecular vibration may
lead to significantly different effects as compared to equilibirum.
This is due to energy absorption effects, that in the present case of
interest is particularly evident in a range of driving frequency for
which the modulation is resonant with the characteristic energies of
the induced non-equilibrium excitations.  

In such a region the effect of modulation competes with the effect of
the interaction imbalance. The average imbalance favours the formation of
a transient superconducting  state with a larger order parameter
extending much beyond the equilibrium critical temperature, while the
time-dependent modulation induces energy absorption into the system.
The latter effect may lead to a depletion of the initial
superconducting state but, in spite of this, it does not preclude the SC order
parameter to become finite above the equilibrium $T_c$ due to the
former, even though with a smaller value with respect to the case in
which the equivalence between the modulated and unmodulated dynamics is established.
Therefore, the signature of a transient SC state  above $T_c$ may
survive also in the region where pairs are resonantly excited, as
shown in Fig.~\ref{fig:above_tc_neqSC} for $\Omega=0.15~\text{eV}$, 
thus realizing a non-trivial dynamical response for which SC is dynamically
extended beyond $T_c$ and not enhanced for $T<T_c$.

In connection with the light-induced transient response in  K$_3$C$_{60}$,
we stress that our results are based on the assumption that the main 
effect of the light pulse is the excitation of the $T_{1u}$ phonon mode that in turn
leads to the discussed interaction imbalance modulation. 
While this is not the only possible outcome of the light excitation
and other types of excitation, \eg of electronic
origin~\cite{michele_excitonC60},  may play an important role, 
the above observations show that the discussed mechanism is a valid source
for a transient superconductivity response above $T_c$ in alkali-doped
fullerides. 

\begin{figure}
  \includegraphics[width=1.\linewidth]{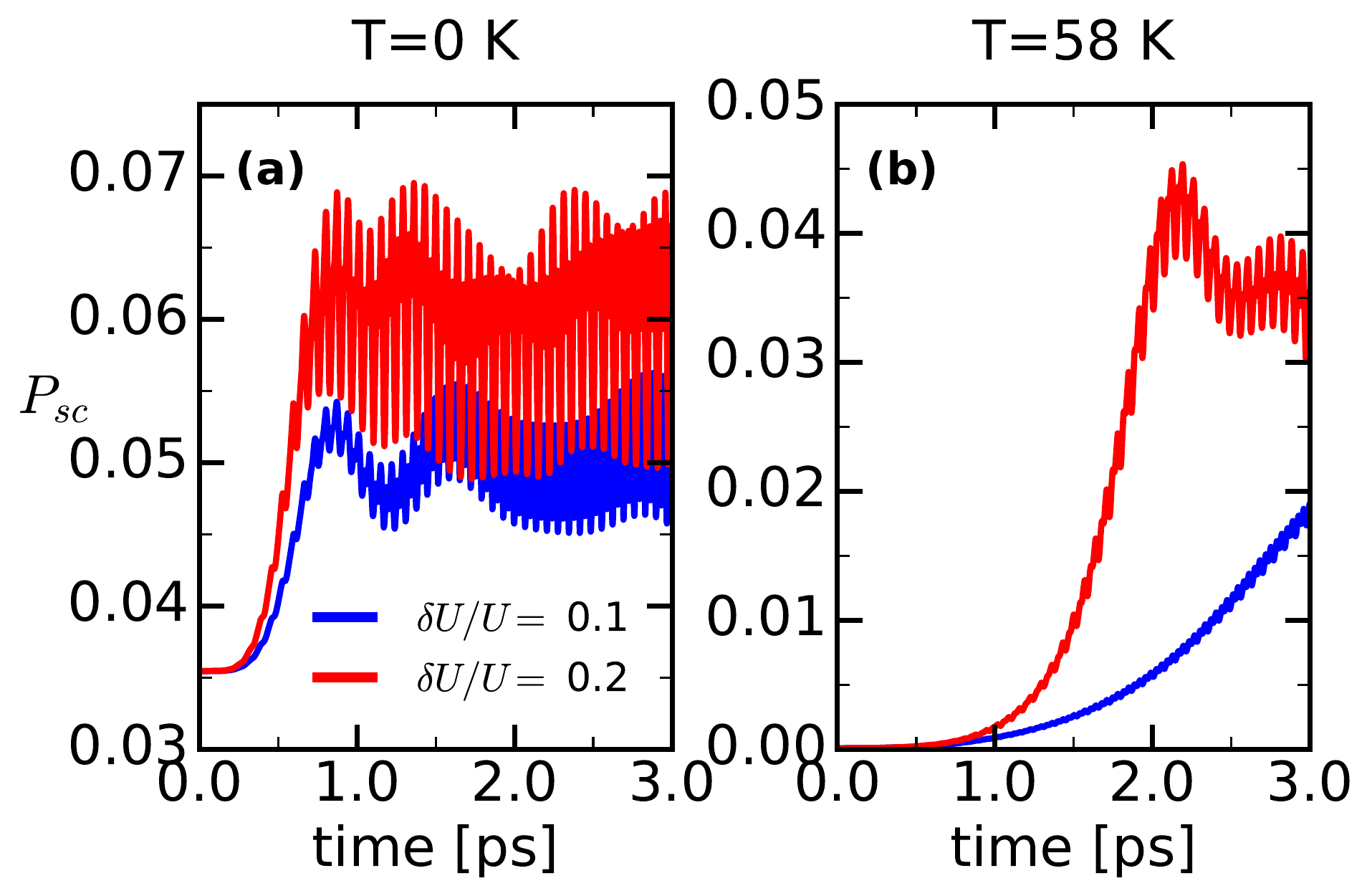}
  \caption{Enhancement of superconductivity as a function of the
    amplitude of the interaction imbalance. $\Omega=0.1875~\text{eV}$.}
  \label{fig:deltaU_dependence}
\end{figure}

Support to the considered mechanism comes from the facts that the
superconducting-like transient 
response is observed for frequencies close to the phononic ones and
for large enough laser fluence ($\gtrsim 1~\text{mJ cm}^{-2}$). 
Under the present assumption, the latter can be understood considering
that a larger laser fluence translates into a larger displacement of
the phonon mode and, therefore, to a larger amplitude of the
interaction imbalance. 
As shown in Fig.~\ref{fig:deltaU_dependence} a larger interaction
amplitude imbalance $\delta U/U=0.2$ leads to an enhancement of the
discussed effects, correctly describing, on a qualitative level, 
the experimental dependence of the transient state on the laser fluence. 
In this respect, it should be noticed that the value of the interaction
imbalance amplitude considered in this minimal model $\delta U/U=0.1$
might be an overestimation of the distorsion that can be induced in a
more realistic description of fullerides~\cite{minjae_dU}.



An important aspect of our theoretical results is that, in a frequency
regime, the dynamical response to the modulation can be different when
the  initial  state is a superconducting state at $T<T_c$ and a normal
state at $T>T_c$. With respect to the available experimental
observations~\cite{mitrano_k3c60,cantaluppi_k3c60} we mention that a
similar effect has been observed in Ref.~\onlinecite{mitrano_k3c60}, where
no enhancement of the transient superconducting gap below $T_c$ has
been reported.

All the above elements emphasize the relevance of the present
mechanism for the description of the transient response of stimulated
fullerides. Despite this, additional evidence is needed in order to fully clarify the
origin of such light-induced transient state. In this respect,
experiments changing the alkali-atom, as \eg 
the Mott insulating compound Cs$_3$C$_{60}$, might be useful to
detect possible transient changes in the electronic correlations
induced by the light pulse.
At the same time an extensive frequency dependence investigation of the 
transient response  below and above $T_c$ would be useful to better 
highlight the possible differences between the two cases, as described
by our results. 



In conclusion we have investigated the non-equilibrium dynamics of a minimal model
describing SC in alkali-doped fullerides subject to the orbital asymmetric periodic
modulation of the local interaction energies. This perturbation results from
the assumption that the effect of the light-pulse is the excitation of
a local vibrational mode and it is known  to enhance SC at equilibrium~\cite{minjae_dU}.
We showed that this leads to the dynamical formation of a superconducting state extending 
beyond the equilibrium critical temperature $T_c$.
Due to the correlated nature of such systems we showed that the transient state, 
while extending SC beyond $T_c$, may show no significant enhancement of the superconducting 
properties below $T_c$ where a SC suppression can even occour.
This captures some non-trivial observations in the light-induced response of \kc 
which validates the mechanism as a possible source of light-induced SC in these
systems.

\section{Acknowledgments.}
We thank Minjae Kim, A.~Cavalleri, M.~Fabrizio, M.~Capone and
H.~Strand for insightful discussions. 
The research leading to these results has received funding from the
European Research Council under the European Union's Seventh Framework
Programme (FP7/2007-2013) ERC Grant Agreement nr. 319286 (Q-MAC).

\section*{Appendix: Gutzwiller approximation for multi-band superconductivity}

In this section we give some details on the time dependent variational
approach used to study the dynamics in the present model. For a
datailed general formulation of the time-dependent Gutzwiller
approximation we refer the reader to
Ref.~\onlinecite{michele_review}. In the following we will discuss the 
extension to the present multi-band superconducting case.

\subsection{General Formulation}

We consider the Hamiltonian defined in section~\ref{sec:model}, which we divide into an 
hopping part $\mathcal{H}_0 = \sum_{\langle i,j \rangle} \sum_{\alpha \beta} t_{i,j}^{\alpha,\beta} 
\cc_{i \alpha} \ca_{j\beta}$ 
and the local interaction $\mathcal{H}_{loc}$ defined in
Eq.~\ref{eq:Hloc}.  Greek indeces include both orbital and spin degrees of freedom.

The dynamics is described starting from the following ansatz for the
time-evolving wavefunction
\begin{equation}
  \ket{\Psi(t)} \equiv \prod_i \mathcal{P}_i (t)
  \left| \Psi_0(t) \right\rangle.
  \label{eq:ansatz}
\end{equation}
$ \left| \Psi_0(t) \right\rangle$ is a single-particle wavefunction
which is meant to describe the dynamics of delocalized quasiparticles
and $\mathcal{P}_i(t)$ are operators acting on the local Hilbert
space defined by the a of Fock states $\ket{\Gamma,i}$.

The variational dynamics is determined by the time-dependent variational 
principle  
\begin{equation}
  \delta \int \braket{\Psi(t)}{i \partial_t - \mathcal{H}}{\Psi(t)} = 0.
\end{equation}
An exact expression for the Lagrangian defining the above variational
principle can be analytically obtained in the limit of infinite lattice
coordination once the following constraints are imposed on the
variational ansatz at each time $t$
\begin{align}
 & \braket{\Psi_0(t)}{\mathcal{P}_i^{\dagger}(t)
  \mathcal{P}_i(t)}{\Psi_0(t)} = 1\\
 &\braket{\Psi_0(t)}{\mathcal{P}_i^{\dagger}(t)
  \mathcal{P}_i(t) \hat{\rho}_i^{N,A}}{\Psi_0(t)} = \braket{\Psi_0(t)}{\hat{\rho}_i^{N,S}}{\Psi_0(t)},
\end{align}
where $\hat{\rho}_i^{N,A}$ are the normal (N) and anomalous (A)
components of the local single particle density matrix defined as
\begin{eqnarray}
  &&\left[ \hat{\rho}_i^N \right]_{\alpha,\beta} = \cc_{i \alpha} \ca_{i
  \beta} \\
  &&\left[ \hat{\rho}_i^A \right]_{\alpha,\beta} = \cc_{i \alpha} \cc_{i
  \beta}.
\end{eqnarray}

A convenient representation of the local projectors is obtained in the
so called mixed original-natural basis representation
\begin{equation}
  \mathcal{P}_i = \sum_{\Gamma,n} \varphi_{i,\Gamma n} (t) \ket{\Gamma,i} \bra{n,i}
\end{equation}  
where $\ket{n,i}$ are the Fock states in the natural basis defined by
a new set of
creation and annhilation operators $\dc_{i,\alpha} \da_{i,\alpha}$ for which
the expectation values of the local single particle density matrix onto the
uncorrelated wavefunction $\ket{\Psi_0(t)}$ is diagonal for the normal component and
zero for the anomalous one
\begin{eqnarray}
  &&\braket{\Psi_0(t)}{\dc_{i \alpha} \da_{i
  \beta} }{\Psi_0(t)}=\delta_{\alpha,\beta} n^0_{i\alpha}(t)\\
&&\braket{\Psi_0(t)}{\dc_{i \alpha} \dc_{i
  \beta} }{\Psi_0(t)}=0 \quad \forall \alpha,\beta.
\end{eqnarray}
The elements $\varphi_{i,\Gamma n} (t)$ are the set of local variational
parameters defining the projectors $\mathcal{P}_i$ and they can
be rewritten in the more convenient form 
\begin{equation}
  \varphi_{i,\Gamma n} (t) = \frac{\Phi_{i,\Gamma n} (t)}{\sqrt{P^{0}_{n,i}(t)}}, 
\end{equation}
where $P^{0}_{n,i}$ are the diagonal occupation probabilities onto the uncorrelated
wavefunction of the local Fock states in the natural basis, expressed
in terms of the variational density matrix $n_{i,\alpha}^0(t)$
\begin{equation}
  \begin{split}
  P^0_{n,i}(t) &= \bra{\Psi_0(t)} \ket{n,i} \bra{n,i} \ket{\Psi_0(t)} \\
  &= \prod_{\alpha} n_{i,\alpha}^0(t)^{n_{\alpha}}(1-n_{i,\alpha}^0(t))^{1-n_{\alpha}}.
  \end{split}
\end{equation}
The matrices $\hat{\Phi}_i$, with elements $\Phi_{i,\Gamma n}$,
contain the set of all the local variational parameters. In the
present spatial homogeneous case all the $\hat{\Phi}_i$ matrices are
equal and we neglect the site index $i$.

With the above definitions the
constraints can be written in terms of  the the local matrices $\hat{\Phi}$
\begin{align}
  & \text{Tr} \left( \Phi^{\dagger}(t) \Phi(t) \right) = 1\\
 & \text{Tr} \left( \Phi^{\dagger}(t) \Phi(t)
    \dc_{i \alpha} \da_{i \beta} \right) 
    = \braket{\Psi_0(t)}{\dc_{i \alpha} \da_{i \beta}}{\Psi_0(t)} =
    \delta_{\alpha \beta} n^0_{\alpha}\\
  & \text{Tr} \left( \Phi^{\dagger}(t) \Phi(t)
    \dc_{i \alpha} \dc_{i \beta} \right) 
    = \braket{\Psi_0(t)}{\dc_{i \alpha} \dc_{i \beta}}{\Psi_0(t)} =0.
\end{align}

Plugging the above definitions into the expression of the Lagrangian
$\mathcal{L}(t) = \braket{\Psi(t)}{i \partial_t
  -\mathcal{H}}{\Psi(t)}$ and imposing the condition $\delta
\int_0^t d\tau \mathcal{L}(\tau) = 0$ the following equations of
motion are obtained~\cite{michele_review}
\begin{eqnarray}
  && i \partial_t \hat{\Phi}(t) = \mathcal{H}_{loc} (t) \hat{\Phi}(t) +
     \braket{\Psi_0(t)}{\frac{\delta \mathcal{H}_{\star} [\hat{\Phi}]}{\delta
     \hat{\Phi}^{\dagger}(t)}}{\Psi_0(t)} \label{eq:iHloc} \\
  && i \partial_t \ket{\Psi_0(t)} = \mathcal{H}_{\star} [\hat{\Phi}] \ket{\Psi_0(t)}.
     \label{eq:iHstar}
\end{eqnarray} 
$\mathcal{H}_{loc}$ is the orginal time-dependent local Hamiltonian and $\mathcal{H}_{\star} [\hat{\Phi}]$ is a
quadratic Hamiltonian which is obtained from the orginial
tight binding Hamiltonian $\mathcal{H}_0$ upone the following
transformation of the fermionic operators
\begin{equation}
  \ca_{i \alpha} \rightarrow \sum_{\beta} R_{\alpha \beta} [\Phi] \da_{i
    \beta} + Q_{\alpha \beta} [\Phi] \dc_{i  \beta}.
  \label{eq:cd_trans}
\end{equation}
The transformation matrices $\mathbf{R} [\hat{\Phi}(t)]$ and $\mathbf{Q}
[\hat{\Phi}(t)]$ entering in Eq.~\ref{eq:cd_trans}
depend on time through the local matrices $\hat{\Phi}(t)$ and their
explicit expressions read
\begin{align}
  &R_{\alpha \beta}[\hat{\Phi}] = \frac{1}{\sqrt{n_{\beta}^0(t)
  (1-n_{\beta}^0(t))}} \text{Tr} \left( \hat{\Phi}^{\dagger}(t)
  \ca_{i\alpha} \hat{\Phi} (t) \dc_{i\beta}   \right) \label{eq:Rdef}\\
  &Q_{\alpha \beta}[\hat{\Phi}] = \frac{1}{\sqrt{n_{\beta}^0(t)
  (1-n_{\beta}^0(t))}} \text{Tr} \left( \hat{\Phi}^{\dagger}(t)
  \ca_{i\alpha} \hat{\Phi} (t) \da_{i\beta}   \right) \label{eq:Qdef}
\end{align}
With the above substitution the quasiparticle Hamiltonian $\mathcal{H}_{\star}$ 
acquires the general matrix form
\begin{equation}
  \mathcal{H}_{\star} = \sum_{\mathbf{k}} \boldsymbol{\Psi}^{\dagger}_{\mathbf{k}} 
  \left(
  \begin{matrix}
    \mathbf{R}^{\dagger} \hat{t}_{\mathbf{k}} \mathbf{R} & \mathbf{R}^{\dagger} \hat{t}_{\mathbf{k}} \mathbf{Q} \\
    \mathbf{Q}^{\dagger} \hat{t}_{\mathbf{k}} \mathbf{R} & \mathbf{Q}^{\dagger} \hat{t}_{\mathbf{k}} \mathbf{Q} 
  \end{matrix}
  \right)
  \boldsymbol{\Psi}_{\mathbf{k}},
\end{equation}
where $\boldsymbol{\Psi}_{\mathbf{k}}^{\dagger}=
\left( \dc_{\mathbf{k},1},~ \dc_{\mathbf{k},2},~ \ldots,~ \dc_{\mathbf{k},N}  
,~\da_{\mathbf{k},1},~ \da_{\mathbf{k},2},~ \ldots,~ \da_{\mathbf{k},N}  
\right)$ and $\hat{t}_{\mathbf{k}}$ is the discrete Fourier transform
of the hopping matrix elements in the Hamiltonian $\mathcal{H}_0.$

The coupled equations of motion \ref{eq:iHloc}-\ref{eq:iHstar} 
describe, respectively, the dynamics of the local degrees of freedom (Eq.~\ref{eq:iHloc})
and of the delocalized quasiparticles (Eq.~\ref{eq:iHstar}). The two dynamics are
coupled in a mean-field way where each degree of freedom provides an
effective field for the other, implying a mutual feedback between the localized and 
delocalized degrees of freedom. This is a big improvement of the present 
method with respect to standard mean-field  (Hartree-Fock) approaches
where only the delocalized quasiparticles are described.
For this reason the method is able to capture correlations effects beyond 
mean-field, as the strongly correlated superconductivity discussed in
the present case. 

\subsection{An explicit implementation}
We now explicity illustrate the implementation of the above general
formulation for the three orbitals model discussed in the main text in
the presence of superconductivity.

The variational wavefunction depends on the choise
of the variational matrix $\hat{\Phi}$ defined in the mixed basis of
original and natural Fock states. In the three orbitals
case the dimension of the local Hilber spase is 64, so that a total number of
64$\times$64=4096 variational parameters  is contained in
$\hat{\Phi}$. 
Such a large number of variational parameter is conveniently reduced
exploiting the symmetries of the Hamiltonian as detailed described in
Ref.~\onlinecite{hugo_lanata}.

The orginal Hamiltonian has a $U(1) \times SU(2) \times O(3)$ symmetry
corresponding to charge conservation and invariance with respect to 
spin and orbital rotations respectively. 
This can be readly appreciated writing the local Hamiltonian in terms of 
the charge operators $N$ and the spin $\mathbf{S}$ and orbital $\mathbf{L}$ 
rotations generators~\cite{demedici_hund_review} 
\begin{equation}
  \mathcal{H}_{loc} = \frac{U-3J_H}{2} \hat{N}\left(\hat{N}-1\right) - 2 J_H \mathbf{S}^2 
  -\frac{J}{2} \mathbf{L}^2 + \frac{5}{2} J_H \hat{N},
\end{equation}
where, restoring the the separation between orbital and spin indeces,
$\hat{N}=\sum_{a \sigma} \cc_{a \sigma} \ca_{a \sigma}$,
$\mathbf{S} = \frac{1}{2} \sum_{a} \sum_{\sigma \sigma'} 
\cc_{a \sigma} \boldsymbol{\tau}_{\sigma \sigma'} \ca_{a \sigma} $ 
and $L_{a} = i \sum_{b c} \sum_{\sigma} \epsilon_{a b c} 
\cc_{b \sigma} \ca_{c \sigma}$, $\boldsymbol{\tau}$ are Pauli matrices and
$\epsilon_{a b c}$ is the Levi-Civita tensor.

In order to describe superconductivity we break the $U(1)$ symmetry
which corresponds to allow for non-zero $\hat{\Phi}$ matrix elements
between states with different number of particles. In such a situation,
the matrix $\mathbf{Q}$ defined in Eq.~\ref{eq:Qdef}  is non zero, so
that the quasiparticle Hamiltonian becomes an effective 
multi-band BCS Hamiltonian.
The removal of the orbital
degeneracy introduced by the imbalanced interaction terms explicitly
breaks the $O(3)$ symmetry so that only the $SU(2)$ symmetry is left.  
We impose the $SU(2)$ symmetry onto the $\hat{\Phi}$ matrix following
the method outlined in Ref.~\onlinecite{hugo_lanata}.
This leads to $N_{\Phi}=429$  independent variational parameters.

The $\hat{\Phi}$ matrix can be expanded onto a basis of
$N_{\Phi}$ matrices satifying 
\begin{equation}
  \text{Tr} \left( \hat{\Phi}^{\dagger}_{k} \hat{\Phi}^{\phantom{\dagger}}_{k'}\right) =
  \delta_{k,k'} \quad k,k'=1,\ldots,N_{\Phi},
\end{equation}
so that all the information about the variational parameters can be
stored in a time-dependent vector  of complex number $\ket{\alpha(t)}$
of dimension $N_{\Phi}$~\cite{hugo_lanata}
\begin{equation}
  \hat{\Phi}(t) = \sum_{k=1}^{N_{\Phi}} \alpha_k(t) \hat{\Phi}_k.
\end{equation}
Next we define the matrices of dimension $N_{\Phi} \times N_{\Phi}$ containg the possible 
combinations of traces over the matrix basis $\hat{\Phi}_k$
\begin{eqnarray}
  && \left[\hat{\rho}^N_{\alpha \beta}\right]_{k,k'} = \text{Tr} \left( \hat{\Phi}^{\dagger}_k \hat{\Phi}_{k'} \cc_{\alpha} \ca_{\beta} \right) \\
  && \left[\hat{\rho}^S_{\alpha \beta}\right]_{k,k'} = \text{Tr} \left( \hat{\Phi}^{\dagger}_k \hat{\Phi}_{k'} \cc_{\alpha} \cc_{\beta} \right) \\
  && \left[ \hat{r}_{\alpha \beta} \right]_{k,k'} = \text{Tr} \left( \hat{\Phi}_k^{\dagger}
  \ca_{i\alpha} \hat{\Phi}_{k'} \dc_{i\beta}   \right)\\
  && \left[ \hat{q}_{\alpha \beta} \right]_{k,k'} = \text{Tr} \left( \hat{\Phi}_k^{\dagger}
  \ca_{i\alpha} \hat{\Phi}_{k'} \da_{i\beta}   \right)\\
  && \left[ \hat{\mathcal{O}}_i \right]_{k,k'} = \text{Tr} \left( \hat{\Phi}_k^{\dagger}
  \mathcal{O}_i \hat{\Phi}_{k'} \right),
\end{eqnarray}
where $\mathcal{O}_i$ represents any local many-body operators.

With the above definitions the constraints and the hopping renormalization matrices 
$\mathbf{R}$ and $\mathbf{Q}$ become
\begin{align}
  & \langle{\alpha(t)} | \alpha(t)\rangle = 1 \label{eq:norm_constr}\\
  & \braket{\alpha(t)}{\hat{\rho}^N_{\alpha,\beta}}{\alpha(t)} = \delta_{\alpha \beta} n^0_{\alpha} (t) \label{eq:densN_constr}\\
  & \braket{\alpha(t)}{\hat{\rho}^A_{\alpha,\beta}}{\alpha(t)} = 0 \quad \forall~\alpha,\beta \label{eq:densA_constr}\\
  & R_{\alpha,\beta}(t) = \frac{ \braket{\alpha(t)}{\hat{r}_{\alpha,\beta}}{\alpha(t)} }{\sqrt{n^0_{\beta}(t)(1-n^0_{\beta}(t))}} \\
  &Q_{\alpha,\beta}(t) = \frac{ \braket{\alpha(t)}{\hat{q}_{\alpha,\beta}}{\alpha(t)} }{\sqrt{n^0_{\beta}(t)(1-n^0_{\beta}(t))}}.
\end{align}
Inserting the above definitions in the evolution of the $\hat{\Phi}$ matrix Eq.~\ref{eq:iHloc},
we obtain the  time evolution in terms of the
$\ket{\alpha(t)}$ vector 
\begin{equation}
  i \partial_t \ket{\alpha(t)} = \hat{\mathcal{H}}_{loc}(t) \ket{\alpha(t)}
  \label{eq:dyn_local_vector}
\end{equation}
$\hat{\mathcal{H}}_{loc}(t)$ is  $N_{\Phi} \times N_{\Phi}$
Hamiltonian defined as follow
\begin{widetext}
\begin{equation}
  \begin{split}
  \left[\widetilde{\mathcal{H}}_{loc} (t)\right]_{k,k'}& =  \left[\mathcal{H}_{loc}(t)\right]_{k,k'} + \\
  & + 
  \sum_{\alpha \beta} \mathcal{D}_{\alpha \beta} \left\lbrace \frac{1}
      {\sqrt{n_{\beta}^0(t) (1-n_{\beta}^0(t))}}\left[\hat{r}_{\alpha,\beta}\right]_{k,k'}
      -\frac{1-2n^0_{\beta}(t)}{2 n_{\beta}^0(t) (1-n_{\beta}^0(t))} \left[\hat{\rho}^N_{\alpha \beta}\right]_{k,k'}
      \right\rbrace + \text{H.c.} + \\
      &+ 
      \sum_{\alpha \beta} \mathcal{S}_{\alpha \beta} \left\lbrace \frac{1}
          {\sqrt{n_{\beta}^0(t) (1-n_{\beta}^0(t))}}\left[\hat{q}_{\alpha,\beta}\right]_{k,k'}
          -\frac{1-2n^0_{\beta}(t)}{2 n_{\beta}^0(t) (1-n_{\beta}^0(t))} \left[\hat{\rho}^A_{\alpha \beta}\right]_{k,k'}
          \right\rbrace + \text{H.c.} .
  \end{split}
\end{equation}
\end{widetext}
The matrices $\mathbf{\mathcal{D}}$ and $\mathbf{\mathcal{S}}$ are defined through the
occupations onto the quasiparticle wavefunction $\ket{\Psi_0(t)}$
\begin{equation}
  \begin{split}
  \mathcal{D}_{\alpha \beta} =\sum_{\mathbf{k} \gamma} &
  \left[ \mathbf{R}^{\dagger} \hat{t}_{\mathbf{k}} \right]_{\gamma
    \alpha}
  \braket{\Psi_0(t)}{\dc_{\mathbf{k}\gamma}\da_{\mathbf{k}\beta}}{\Psi_0(t)}
  \\
  +&
  \left[ \mathbf{Q}^{\dagger} \hat{t}_{\mathbf{k}} \right]_{\gamma
    \alpha}
  \braket{\Psi_0(t)}{\da_{\mathbf{k}\gamma}\da_{\mathbf{k}\beta}}{\Psi_0(t)}
  \end{split}
\end{equation}
\begin{equation}
\begin{split}
\mathcal{S}_{\alpha \beta} =\sum_{\mathbf{k} \gamma} &
  \left[ \mathbf{R}^{\dagger} \hat{t}_{\mathbf{k}} \right]_{\gamma
    \alpha}
  \braket{\Psi_0(t)}{\dc_{\mathbf{k}\gamma}\dc_{\mathbf{k}\beta}}{\Psi_0(t)}
  \\
  +& \left[ \mathbf{Q}^{\dagger} \hat{t}_{\mathbf{k}} \right]_{\gamma
    \alpha}
  \braket{\Psi_0(t)}{\da_{\mathbf{k}\gamma}\dc_{\mathbf{k}\beta}}{\Psi_0(t)}
\end{split}
\end{equation}

The dynamics described by Eq.~\ref{eq:dyn_local_vector} is coupled to the dynamics of the 
wavefunction $\ket{\Psi_0(t)}$ (Eq.~\ref{eq:iHstar}) which can be fully expressed 
through the dynamics of the occupations $\Delta_{\mathbf{k}}^{\alpha,\beta}(t) = 
\braket{\Psi_0(t)}{\dc_{\mathbf{k} \alpha} \da_{\mathbf{k} \beta}}{\Psi_0(t)}
$ and $\Gamma_{\mathbf{k}}^{\alpha,\beta}(t) = 
\braket{\Psi_0(t)}{\dc_{\mathbf{k} \alpha} \dc_{\mathbf{k} \beta}}{\Psi_0(t)}$

\begin{eqnarray}
  &&i \partial_{t} \Delta_{\mathbf{k}}^{\alpha,\beta}(t) = \braket{\Psi_0(t)}{[\dc_{\mathbf{k} \alpha} \da_{\mathbf{k} \beta},\mathcal{H}_{\star}]}{\Psi_0(t)} \label{eq:normal_slater}\\
  &&i \partial_{t} \Gamma_{\mathbf{k}}^{\alpha,\beta}(t) = \braket{\Psi_0(t)}{[\dc_{\mathbf{k} \alpha} \dc_{\mathbf{k} \beta},\mathcal{H}_{\star}]}{\Psi_0(t)}\label{eq:anomalous_slater}
\end{eqnarray}
The system of coupled Eqs.~\ref{eq:dyn_local_vector},~\ref{eq:normal_slater} and \ref{eq:anomalous_slater}
fully describe the dynamics within the variational ansatz Eq.~\ref{eq:ansatz} in a lattice
with infinite coordination number.

The number of coupled differential equations is further reduced by the
fact that in the present case the hopping matrix is diagonal in both
spin and orbital indeces and  we consider only the possibility of
intra-orbital pairing.
This means that $\Delta_{\mathbf{k}}^{\alpha,\beta} = \delta_{\alpha,\beta} \Delta_{\mathbf{k}}^{\alpha}$ and 
$\Gamma_{\mathbf{k}}^{\alpha,\beta} = \delta_{a,b} (1-\delta_{\sigma \sigma'}) \Gamma_{\mathbf{k}}^{a \sigma, a \sigma'}$,
where in the last case we separated orbital $(a,b)$ and spin $(\sigma,\sigma')$ degrees of freedom.
Moreover, the matrix $\mathbf{R}$ is diagonal in both orbital and spin indeces, while the matrix $\mathbf{Q}$ 
is diagonal in the orbital index and couples only states with opposite spin.

The dynamics is unitary and preserves the normalization constraints Eq.~\ref{eq:norm_constr} and the 
density constraint Eq.~\ref{eq:densN_constr} and they need to be enforced only at equilibrium.
On the contrary, we found that the constraint on the anomalous density is no more conserved by the unitary dynamics
as the particle-hole symmetry is lifted by the imbalanced interaction. In that case we introduce a
set of time-dependent Lagrange multipliers enforcing the constraints at each time step.

The equations of motions are solved using the explicit  4th-order Runge-Kutta method with a 
time discretization $\delta t = 0.01$ starting from the variational estimation of the 
equilibrium ground state. The latter is obtained from the stationary limit of the equations of
motion
\begin{eqnarray}
  &&\Lambda \ket{\alpha} = \widetilde{\mathcal{H}}_{loc} \ket{\alpha}\\
  &&E_{\star} \ket{\Psi_0} = \mathcal{H}_{\star} \ket{\Psi_0}.
\end{eqnarray}
This corresponds to finding the ground state of a non-linear eigenvalue 
problem with appropriate Lagrange parameters enforcing the constraints
at $t=0$.
This can be done recursively at fixed value of the variational density
matrix $n^0$.~\cite{hugo_lanata}
Therefore, a full minimization of the obtained ground state energy with respect to $n^0$ gives 
the variational estimation of the ground state energy.

\subsection{FINITE TEMPERATURE}
The finite temperature extension closely follows the zero-temperature one 
where the variational ansatz for the time evolving ground state is replaced by
the ansatz for the time dependent density matrix
\begin{equation}
  \begin{split}
  \rho(t) = &\sum_n p_n \ket{\Psi_{n}(t)} \bra{\Psi_{n}(t)}   = \\
  &\sum_n p_n \mathcal{P}(t) \ket{\Psi_{0,n}(t)} \bra{\Psi_{0,n}(t)}\mathcal{P}^{\dagger}(t) = \\
  &\mathcal{P}(t) \rho_{\star} (t) \mathcal{P}^{\dagger}(t),
  \end{split}
  \label{eq:density_matrix_ansatz}
\end{equation}
namely the Gutzwiller projection is done on each uncorrelated state $\ket{\Psi_{0,n}(t)}$
describing an uncorrelated density matrix
$\rho_{\star}(t) = \sum_{n} p_{n} \ket{\Psi_{0,n}(t)} \bra{\Psi_{0,n}(t)}$ 
through the distribution $p_n$. 
The Dirac-Frenkel extension of the time-dependent variational principle
reads 
\begin{equation}
  \delta \int \sum_n p_n \braket{\Psi_n(t)}{i \partial_t -\mathcal{H}}{\Psi_n(t)} = 0.
\end{equation}
With the same definitions of projectors and constraints as in the previous section
 the equation of motions are equivalent to the zero-temperature case with 
the averages onto the uncorrelated wavefunction replaced by traces over the uncorrelated
density matrix
\begin{align}
  & i \partial_t \hat{\Phi}(t) = \mathcal{H}_{loc} \hat{\Phi}(t) +
  \sum_n p_n \braket{\Psi_{0,n}(t)}{\frac{\delta \mathcal{H}_{\star} [\hat{\Phi}]}{\delta
     \hat{\Phi}^{\dagger}(t)}}{\Psi_{0,n}(t)} \label{eq:iHloc_T} \\
  & i \partial_t \ket{\Psi_{0,n}(t)} = \mathcal{H}_{\star} [\hat{\Phi}] \ket{\Psi_{0,n}(t)}
     \label{eq:iHstar_T}.
\end{align} 
A practical solution of Eqs.~\ref{eq:iHloc_T}-\ref{eq:iHstar_T} is obtained starting the initial
estimation of the finite temperature expectation values for the quasiparticle occupations
$\Delta_{\mathbf{k}}^{\alpha,\beta}(t) = \text{Tr}\left( \rho_{\star}(t) \dc_{\mathbf{k} \alpha} \da_{\mathbf{k} \beta} \right)$
and
$\Gamma_{\mathbf{k}}^{\alpha,\beta}(t) = \text{Tr}\left( \rho_{\star}(t) \dc_{\mathbf{k} \alpha} \dc_{\mathbf{k} \beta} \right)$
leading to the following equations of motions
\begin{eqnarray}
  &&i \partial_{t} \Delta_{\mathbf{k}}^{\alpha,\beta}(t) = 
  \text{Tr} \left( \rho_{\star}(t) [\dc_{\mathbf{k} \alpha} \da_{\mathbf{k} \beta},\mathcal{H}_{\star}] \right)\\
  &&i \partial_{t} \Gamma_{\mathbf{k}}^{\alpha,\beta}(t) = 
  \text{Tr} \left( \rho_{\star}(t) [\dc_{\mathbf{k} \alpha} \dc_{\mathbf{k} \beta},\mathcal{H}_{\star}] \right).
\end{eqnarray}
which can be written only in terms of combinations of $\Delta_{\mathbf{k}}^{\alpha,\beta}$ and
$\Gamma_{\mathbf{k}}^{\alpha,\beta}$. Therefore the explicit time evolution for each $\ket{\Psi_{0,n}(t)}$ 
Eq.~\ref{eq:iHstar_T} is, in practice, not needed and the distribution $p_n$ is defined once for all 
by the initial variational estimation of the thermal state. 

The initial thermal state is obtained by minimizaing the variational estimation of the free energy.~\cite{michele_gzT}
In doing so, for a simplification of the minimization procedure, we neglect the contribution coming from the
entropy of the local degrees of freedom. 
This is a reasonable approximation since we focus on the weak coupling side of the superconducting
dome far from the Mott transition, where the contribution of the entropy of the 
local degrees of freedom is small. Its inclusion would lead to a weak renormalization of the equilibrium 
transition temperature, but no qualitative difference in the dynamics is expected with respect to what
reported in the main text.

\bibliography{biblio_neq_fullerides}

\end{document}